\documentclass[]{pasj01}
\usepackage{lineno}

\begin{document} 
\Received{03-Jun-2020}
\Accepted{30-Oct-2020}

\title{The relation between quasars' optical spectra and variability}


\author{Shumpei Nagoshi\altaffilmark{1,}$^{*}$}%
\altaffiltext{1}{Department of Astronomy, Kyoto University, Kyoto 606-8502, Japan}
\email{shumpei@kusastro.kyoto-u.ac.jp}

\author{Fumihide Iwamuro \altaffilmark{1}}


\maketitle

\begin{abstract}
Brightness variation is an essential feature of quasars, but its mechanism and relationship to other physical quantities are not understood well. We aimed to find the relationship between the optical variability and spectral features to reveal the regularity behind the random variation. It is known that quasar's $\rm{FeII}/{\rm H\beta}$ flux ratio and equivalent width of ${\rm [OIII]5007}$ are negatively correlated, called Eigenvector~1. In this work, we visualized the relationship between the position on this Eigenvector 1 (EV1) plane and how they had changed their brightness after $\sim$~10 years. We conducted three analyses using different quasar sample each. The first analysis showed the relation between their distributions on the EV1 plane and how much they had changed brightness, using 13,438 Sloan Digital Sky Survey quasars. This result shows how brightness changes later are clearly related to the position on the EV1 plane. In the second analysis, we plotted the sources reported as Changing-Look Quasars (or Changing-State Quasars) on the EV1 plane. This result shows that the position on the EV1 plane corresponds activity level of each source, the bright or dim state of them are distributed on the opposite sides divided by the typical quasar distribution. In the third analysis, we examined the transition vectors on the EV1 plane using sources with multiple-epoch spectra. This result shows that the brightening and dimming sources move on the similar path and they turn into the position corresponding to the opposite activity level. We also found this trend is opposite to the empirical rule that $R_{\rm{FeII}}$ positively correlated with the Eddington ratio, which has been proposed based on the trends of a large number of quasars. From all these analyses, it is indicated that quasars tend to oscillate between both sides of the distribution ridge on the EV1 plane, each of them corresponds to a dim state and a bright state. This trend in optical variation suggests that significant brightness changes, such as Changing-Look quasars, are expected to repeat.
\end{abstract}

\section{Introduction} \label{sec:intro}
One of the significant characteristics of quasars is their brightness variability, which we can observe at all wavelengths (e.g., \cite{1963ApJ...138...30M, 1997iagn.book.....P}). Their amplitudes and timescales are diverse. In particular, the recent accumulation of survey data has led to studies of the long-term variability at various wavelengths. For example, (optical) Changing-Look/State Quasars (Although the exact definitions are different from Changing-Look Quasars and Changing-State Quasars, we unify the terminology to CLQ hereafter because we refer them in a same context), which show significant flux variations ($> 30\%$) in their broad emission lines, have been discovered in recent years \citep{2015ApJ...800..144L, 2016MNRAS.457..389M, 2019ApJ...874....8M, 2016ApJ...826..188R, 2017ApJ...835..144G, 2018ApJ...858...49W, 2018MNRAS.480.3898N, 2020MNRAS.498.2339R, 2020MNRAS.491.4925G, 2021PASJ...73..122N, 2021PASJ...73..596W}. In X-ray, quasars that the hydrogen column density of the same source changes drastically are known and called with the same name, (X-ray) Changing-look Quasar (e.g., \cite{2009ApJ...696..160R}). In addition, in the radio band, transit from the radio-quiet quasar to the radio-loud quasar has been reported \citep{2020ApJ...905...74N}. The mechanisms of these phenomena are still under debate, but many papers infer that they are expected to relate to the change in the mass accretion rate. In case of significant variation, such as CLQs, the result of X-ray observation \citep{2016A&A...593L...9H} and optical polarization observations \citep{2017A&A...604L...3H} implies that they also undergo rapid changes in intrinsic accretion power. In this study, to conduct a statistical investigation of the status of quasars' accretion disk, we focused on the optical variation because a large amount of data linked to the accretion disk is available.

While quasars' light curves typically look random, the relation between luminosity and other physical quantities has been studied to understand their diverse variation pattern. Some studies have found that the smaller the Eddington ratio are, the more significant variable they are (e.g., \cite{2008MNRAS.383.1232W, 2010ApJ...721.1014M}). One interpretation of this correlation is that the inefficient mass accretion invokes intermittent avalanche-like accretion, resulting in the more significant variability \citep{1995PASJ...47..617T, 1998ApJ...504..671K, 2020ApJ...903...54T}. However, in the case of this avalanche-induced variability, brightening and dimming have the same probability of occurrence if the observing period of the light curve is longer than a few years \citep{1998ApJ...504..671K}. On the other hand, studies of the structure-function of Sloan Digital Sky Survey (SDSS) light curves have shown that the probability of dimming is higher for long-term variability \citep{2005AJ....129..615D}, which cannot be explained by avalanche picture alone. In other words, we still poorly understand the relation between physical quantities and variability.

This study aims to find relations between random variability (in particular, variability with large amplitude and timescale) and the properties of quasars. One of the properties that characterize quasars' spectra is called Eigenvector 1 (EV1), which is a correlation that oxygen and iron emission line intensities are anti-correlated, found by principal component analysis of spectra \citep{1992ApJS...80..109B}. Since EV1 is one of the few regularities that characterize the diverse quasars' spectra, we investigate the relation between locations on the EV1 plane and how they vary later; the EV1 plane here refers to the values of equivalent width (ratio) of $EW({\rm FeII})$/$EW({\rm H\beta})$ ($\equiv R_{\rm Fe_{II}}$) and $\rm{log_{10}}EW({\rm [OIII]5007})$.

Through this investigation of the light curves, we can also expect to gain new insights into the nature of EV1. The physical interpretation of EV1 is limited because it was found as an empirical rule by statistical methods. Some studies suggested that the Eddington ratio is one of the origins of this anti-correlation because the emission lines are mainly photoionized by the UV continuum, which depends on the state of the accretion disk represented by the Eddington ratio \citep{1992ApJS...80..109B, 2000ApJ...536L...5S, 2014Natur.513..210S}. However, the correlation between the Eddington ratio and EV1 has not been confirmed in individual object. To examine the contribution of the Eddington ratio to the EV1 based on the variability, we investigated the transition vectors (vectors showing where it moved from and to on the EV1 plane) on the EV1 plane associated with the changing luminosity of each object.

This paper consists of three sections of analyses, and the structure of this paper is as follows. Section 2 describes the methods and results for each of the three analyses. Section 3 discusses the results of them, as well as observational predictions and limitations of this study. The cosmological parameters used in this study are consistently $H_0=70\ \rm{km/s/Mpc} $, $\Omega_m = 0.3 $ (These values are same as the \citet{2011ApJS..194...45S} catalog we used).

\section{Method and Result} \label{sec:method}

\subsection{Sample}
\label{sec:sample}
We visualize how positions on the EV1 plane and brightness variation relate from three aspects. In the first analysis, we show general trends of the relation between the position on the EV1 plane and how they change brightness later. Second, we show the correspondence relation between the position on the EV1 plane and the accretion state of the quasars. Finally, we show how each quasar moves on the EV1 plane due to its brightness variation. Each of the three analyses uses a different sample. First of all, we summarize the three samples here.

\begin{description}
    \item[Sample~1] 
    : We started from the quasars included in the \citet{2011ApJS..194...45S} catalog (hereafter referred to as S11). To obtain values of $\rm{FeII}$, H$\beta$, and ${\rm [OIII]5007}$, we picked up whose redshift is less than 0.8 (19,480 objects). Then, we excluded quasars with any equivalent width value of used emission lines below its error (remaining 15,446 objects). To estimate their brightness when the samples have these catalog values, we narrowed it down to whose spectroscopic observation period was less than two years from the observation period of the SDSS DR7 photometry (remaining 13,438 objects). As a result, 105,783 quasars in S11 were reduced to 13,438 quasars as Sample~1.

    \item[Sample~2] 
    : This sample consists of CLQs, which are known to be a highly variable quasar population, to investigate the relationship between their position on the EV1 plane and the activity state. Among the sources in S11, 76 quasars that were later reported in \citet{2019ApJ...874....8M, 2016MNRAS.457..389M, 2020MNRAS.491.4925G, 2018ApJ...862..109Y, 2018ApJ...864...27S} as CLQs were listed as sample~2 (Table~1).
    
    \item[Sample~3] 
    : To investigate how each quasar moves on the EV1 plane associated with its optical variation, we collected quasars with multiple spectra as Sample~3. From the 750,414 quasars in the Sloan Digital Sky Survey Data Release 16th Quasar catalog (\citet{2020ApJS..249....3A}; SDSS DR16Q), we narrowed down the list to those with redshifts below 0.8 and multiple spectra available (8142 objects). We fitted their spectra to obtain the position on the EV1 plane. Finally, 2,839 sources remained as Sample~3, in which all emission lines of multiple spectra were fitted with errors less than 10\% of their intensities.
\end{description}

\subsection{Analysis 1: the distribution on the EV1 plane and the subsequent variation}

\begin{figure*}
\begin{center}
\includegraphics[width=160mm]{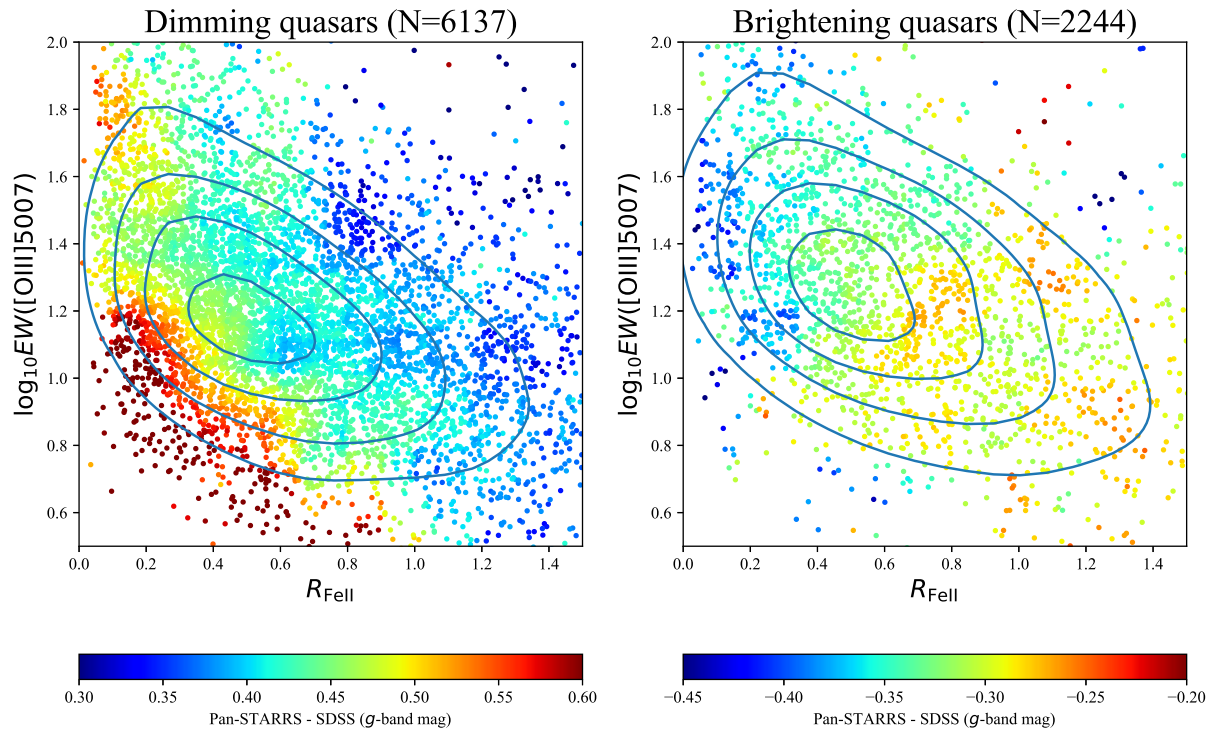}
\end{center}
\caption{Distribution of dimming/brightening quasars on the EV1 plane. The location of each point is based on the S11 catalog with the subsequent brightness variation indicated by the symbol color. The contour lines show the distribution density of each quasar, estimated using Gaussian kernel density estimation. From the outside, they represent the 90, 70, 50, and 30 percent probability of existence. The color bar is the magnitude difference between Pan-STARRS and SDSS in the $g$-band, which is approximately the amount of brightness variation 10 years after the spectra were acquired. The color of each point is averaged for the points within $\pm 0.1$ in the x and y-axis direction.\label{fig:shen_variation}}
\end{figure*}

We visualized the general relationship between the position on the EV1 plane and the brightness variations in $\sim$10 years from the data of Sample~1 in Figure~\ref{fig:shen_variation}. Each position of dot on the Figure~\ref{fig:shen_variation} correspond to the values of S11 (we defined $R_{\rm{Fe}}$ as equivalent width of $\rm{FeII}$ in the range from 4434~$\rm{\AA}$ to 4684~$\rm{\AA}$ divided with the value of equivalent width of $\rm{H\beta}$). The subsequent optical variations were obtained from the differences of the $g$-band magnitudes between SDSS DR7 \citep{2009ApJS..182..543A} and Pan-STARRS DR1 \citep{2016arXiv161205560C} catalogs. The SDSS photometric and spectroscopic observations of the target objects were made between 1999 and 2005 (the median year is 2002), while the Pan-STARRS observations were made between 2009 and 2014 (the median year is 2012); intervals of them are about ten years. Since the photometric system of Pan-STARRS is almost the same as that of SDSS with the systematic differences of less than 0.02~mag \citep{2012ApJ...750...99T}, so we interpret the $g$-band photometric differences to represent how much brightness in the $g$-band magnitudes has changed in about ten years after the spectra were obtained.

Figure~\ref{fig:shen_variation} shows the distribution on the EV1 plane with the amount of the magnitude variation for each group of brightening and dimming sources (``a brightening/dimming source'' means the quasar that was brightening/dimming when their spectra were acquired). We plot brightening and dimming sources in different groups because their numbers are about three times different. There are 2,244 objects with magnitude increases of more than 0.15 mag, 6,136 with decreases of more than 0.15 mag, and 5,042 with variations less than 0.15 mag. The color of the symbols are smoothed using the surrounding points (the color represents the averaged magnitude variation of the points within the box of $\Delta R_{\rm{FeII}} = 0.1$ and $\Delta {\rm log_{10}}EW({\rm [OIII]5007}) = 0.1$ around each point).  

From the color pattern on Figure~\ref{fig:shen_variation}, we can see that some relation exists between the position on the EV1 plane and how their brightness change later. In the left panel of Figure~\ref{fig:shen_variation}, we can see that quasars located on the lower left side of the EV1 plane tend to dim more significantly. In the right panel of Figure~\ref{fig:shen_variation}, there looks to be a weak trend that quasars on the left side of the EV1 plane tend to brighten more significantly, but the sample size is smaller than dimming sources. When we compare the left panel and the right panel of Figure~\ref{fig:shen_variation}, we can see that dimming sources distribute a little lower than brightening sources on the EV1 plane. Overall, it is expected that there is a correspondence between the position on the EV1 plane and the activity change of the quasars.

\subsection{Analysis 2: Distribution of CLQs on the EV1 plane}

\begin{figure*}[ht!]
\includegraphics[width=160mm]{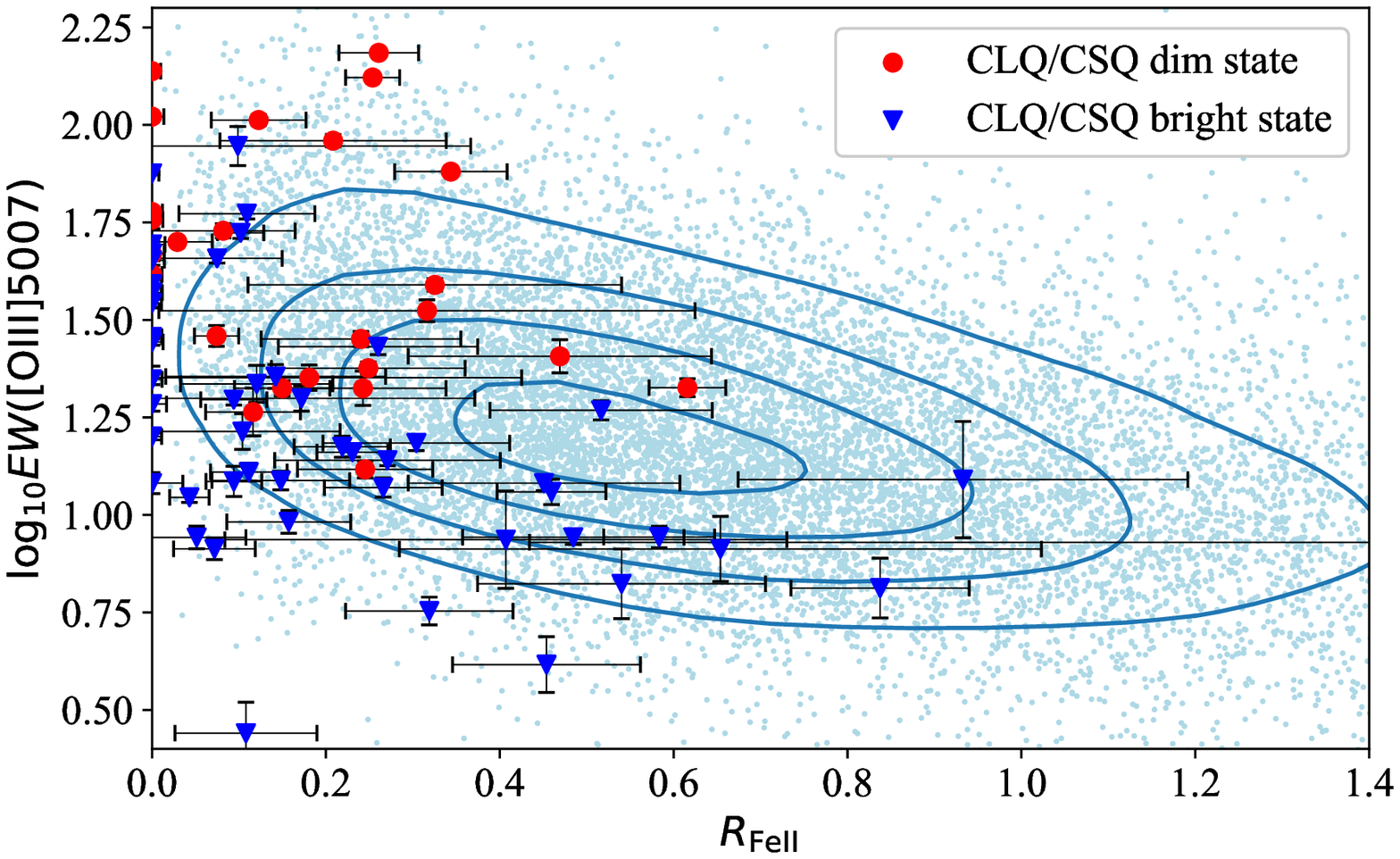}
\caption{
Location of CLQs listed in Table~\ref{clqtable} on the EV1 plane. The red circles represent CLQs in the dim state, and the blue triangles represent CLQs in the bright state. The light blue dots represent other quasars in S11 with redshifts below 0.8. The distribution densities are shown as contours using Gaussian kernel density estimation. From the outside, each contour represents the 90th, 70th, 50th, and 30th percentiles.}
\label{fig:clq_ev1}
\end{figure*}

In this analysis, to confirm the correspondence relation between the position on the EV1 plane and the activity level of the quasars, we investigated the distribution of the CLQs whose activity states are known (Figure~\ref{fig:clq_ev1}). We used 76 CLQs in Table~\ref{clqtable} referred to the papers listed in it. In Figure~\ref{fig:clq_ev1}, we plot CLQs in different markers corresponding to their activity states; red circles represent the bright state, and blue triangles represent the dim state. From Figure~\ref{fig:clq_ev1}, we can see that the bright state and the dim state of CLQs are separated by the ridge of the distribution (the region of high distribution density contributing to the negative correlation) of other S11 quasars. In other words, the lower position on the EV1 plane corresponds to a bright (active) state, and the upper position corresponds to a dim (quiet) state for each quasar.

\begin{longtable}{*{7}{l}}

\caption{
List of Changing-Look/State Quasars reported by 2020 and included in S11, ordered by RA \citep{2019ApJ...874....8M, 2016MNRAS.457..389M, 2020MNRAS.491.4925G, 2018ApJ...862..109Y, 2018ApJ...864...27S}. The SDSS Name, Redshift, $R_{\rm{FeII}}$, and log(EW) are referred to as the values from the S11. The amount of variation in $g$-band magnitudes from SDSS to Pan-STARRS is shown in $\Delta g$ mag.
\label{clqtable}}

\hline
\multicolumn{7}{l}{} \\
SDSS Name & State & Reference & Redshift & $R_{\rm{FeII}}$ & ${\rm log_{10}}EW({\rm [OIII])}$ & $\Delta g$ mag \\
\hline
\endfirsthead
\hline
SDSS Name & State & Reference & Redshift & $R_{\rm{FeII}}$ & ${\rm log_{10}}EW({\rm [OIII])}$ & $\Delta g$ mag \\
\hline
\endhead
\hline
\endfoot
\hline
\multicolumn{7}{l}{} \\
\hline
\endlastfoot

000904.54$-$103428.7 & Bright state & MacLeod et al. 2018 & 0.241 & $0.000 \pm{0.014}$ & $1.670 \pm{0.012}$ & $+$1.37 \\
002311.06$+$003517.5 & Dim State & MacLeod et al. 2016 & 0.422 & $0.249 \pm{0.111}$ & $1.376 \pm{0.036}$ & $-$0.88 \\
011919.27$-$093721.7 & Bright state & Graham et al. 2019 & 0.383 & $0.120 \pm{0.088}$ & $1.336 \pm{0.109}$ & $+$0.05 \\
015957.64$+$003310.4 & Bright state & MacLeod et al. 2016 & 0.312 & $0.000 \pm{0.035}$ &$ 1.082 \pm{0.062}$ & $+$0.08 \\
022014.57$-$072859.2 & Dim State & Graham et al. 2019 & 0.213 & $0.150 \pm{0.055}$ & $1.325 \pm{0.046}$ & $-$1.42 \\
022556.07$+$003026.7 & Bright state & MacLeod et al. 2016 & 0.504 & $0.654 \pm{0.369}$ & $0.913 \pm{0.190}$ & $+$0.86 \\
022652.24$-$003916.5 & Bright state & MacLeod et al. 2016 & 0.625 & $0.407 \pm{0.323}$ & $0.937 \pm{0.280}$ & $+$1.30 \\
025505.68$+$002522.9 & Bright state & Graham et al. 2019 & 0.353 & $0.266 \pm{0.068}$ & $1.070 \pm{0.057}$ & $+$2.44 \\
074511.98$+$380911.3 & Bright state & MacLeod et al. 2018 & 0.237 & $0.485 \pm{0.127}$ & $0.943 \pm{0.043}$ & $+$1.09 \\
075440.32$+$324105.2 & Dim State & Graham et al. 2019 & 0.411 & $0.344 \pm{0.065}$ & $1.880 \pm{0.018}$ & $-$0.34 \\
081425.89$+$294115.6 & Bright state & Graham et al. 2019 & 0.374 & $0.000 \pm{0.012}$ & $1.570 \pm{0.037}$ & $+$1.41 \\
081632.12$+$404804.8 & Bright state & Graham et al. 2019 & 0.701 & $0.104 \pm{0.112}$ & $1.214 \pm{0.105}$ & $+$1.23 \\
082033.30$+$382419.7 & Bright state & Graham et al. 2019 & 0.648 & $0.142 \pm{0.126}$ & $1.355 \pm{0.070}$ & $+$0.41 \\
082930.59$+$272822.7 & Bright state & Graham et al. 2019 & 0.321 & $0.000 \pm{0.014}$ & $1.653 \pm{0.030}$ & $+$0.22 \\
083225.34$+$370736.2 & Dim State & Graham et al. 2019 & 0.092 & $0.000 \pm{0.011}$ & $1.617 \pm{0.027}$ & $-$0.15 \\
083236.28$+$044505.9 & Dim State & Graham et al. 2019 & 0.292 & $0.000 \pm{0.012}$ & $1.777 \pm{0.030}$ & $-$1.27 \\
084716.03$+$373218.0 & Dim State & Graham et al. 2019 & 0.454 & $0.261 \pm{0.046}$ & $2.185 \pm{0.016}$ & $-$0.33 \\
084957.78$+$274729.0 & Bright state & Yang et al. 2018 & 0.299 & $0.271 \pm{0.130}$ & $1.140 \pm{0.033}$ & $+$0.53 \\
091357.26$+$052230.7 & Bright state & Graham et al. 2019 & 0.346 & $0.111 \pm{0.044}$ & $1.110 \pm{0.052}$ & $+$1.15 \\
092441.08$+$284730.3 & Bright state & Graham et al. 2019 & 0.464 & $0.109 \pm{0.078}$ & $1.772 \pm{0.031}$ & $-$0.07 \\
092736.79$+$153823.1 & Bright state & Graham et al. 2019 & 0.555 & $0.099 \pm{0.268}$ & $1.946 \pm{0.115}$ & $+$0.15 \\
092836.78$+$474245.8 & Bright state & Graham et al. 2019 & 0.830 & $0.318 \pm{0.200}$ & $1.916 \pm{0.185}$ & $+$0.86 \\
093017.70$+$470721.0 & Dim State & Graham et al. 2019 & 0.160 & $0.000 \pm{0.008}$ & $1.669 \pm{0.017}$ & $-$0.79 \\
094620.86$+$334746.9 & Dim State & Graham et al. 2019 & 0.239 & $0.616 \pm{0.044}$ & $1.326 \pm{0.053}$ & $-$1.08 \\
095750.03$+$530104.7 & Dim State & Graham et al. 2019 & 0.437 & $0.208 \pm{0.130}$ & $1.960 \pm{0.036}$ & $-$1.39 \\
100220.17$+$450927.3 & Bright state & MacLeod et al. 2016 & 0.400 & $0.454 \pm{0.108}$ & $0.616 \pm{0.163}$ & $+$1.14 \\
100256.21$+$475027.7 & Dim State & Graham et al. 2019 & 0.391 & $0.000 \pm{0.014}$ & $2.021 \pm{0.015}$ & $-$0.59 \\
100343.23$+$512610.8 & Dim State & Graham et al. 2019 & 0.431 & $0.469 \pm{0.174}$ & $1.407 \pm{0.098}$ & $-$0.65 \\
101152.98$+$544206.4 & Bright state & Runnoe et al. 2016 & 0.246 & $0.837 \pm{0.103}$ & $0.812 \pm{0.175}$ & $+$1.40 \\
102152.34$+$464515.6 & Bright state & MacLeod et al. 2016 & 0.204 & $0.230 \pm{0.041}$ & $1.161 \pm{0.029}$ & $+$1.25 \\
102613.90$+$523751.2 & Dim State & Graham et al. 2019 & 0.259 & $0.029 \pm{0.040}$ & $1.700 \pm{0.023}$ & $+$0.13 \\
102817.67$+$211507.4 & Dim State & Graham et al. 2019 & 0.365 & $0.123 \pm{0.054}$ & $2.012 \pm{0.030}$ & $-$0.55 \\
104254.79$+$253713.6 & Dim State & Graham et al. 2019 & 0.603 & $0.000 \pm{0.010}$ & $2.138 \pm{0.024}$ & $-$0.64 \\
105203.55$+$151929.5 & Bright state & Stern et al. 2018 & 0.303 & $0.000 \pm{0.017}$ & $1.284 \pm{0.041}$ & $+$0.82 \\
105553.51$+$563434.4 & Dim State & MacLeod et al. 2018 & 0.322 & $0.325 \pm{0.215}$ & $1.590 \pm{0.037}$ & $-$1.03 \\
110455.17$+$011856.6 & Bright state & Yang et al. 2018 & 0.575 & $0.933 \pm{0.259}$ & $1.091 \pm{0.330}$ & $+$2.10 \\
111329.68$+$531338.7 & Bright state & MacLeod et al. 2018 & 0.239 & $0.148 \pm{0.079}$ & $1.088 \pm{0.055}$ & $+$1.25 \\
111617.80$+$251035.7 & Bright state & Graham et al. 2019 & 0.534 & $0.000 \pm{0.011}$ & $1.348 \pm{0.032}$ & $-$0.25 \\
113111.14$+$373709.1 & Bright state & Graham et al. 2019 & 0.448 & $0.000 \pm{0.011}$ & $1.543 \pm{0.029}$ & $+$1.55 \\
113706.84$+$013947.9 & Bright state & Graham et al. 2019 & 0.193 & $0.043 \pm{0.023}$ & $1.045 \pm{0.029}$ & $+$0.30 \\
114408.90$+$424357.5 & Bright state & Graham et al. 2019 & 0.272 & $0.000 \pm{0.013}$ & $1.443 \pm{0.019}$ & $+$0.48 \\
115039.32$+$363258.4 & Bright state & Yang et al. 2018 & 0.340 & $0.260 \pm{0.115}$ & $1.431 \pm{0.047}$ & $+$1.24 \\
115227.48$+$320959.4 & Bright state & Yang et al. 2018 & 0.374 & $0.075 \pm{0.075}$ & $1.658 \pm{0.027}$ & $+$1.30 \\
120130.63$+$494048.9 & Dim State & Graham et al. 2019 & 0.392 & $0.240 \pm{0.115}$ & $1.451 \pm{0.045}$ & $-$0.79 \\
120442.10$+$275411.7 & Dim State & Graham et al. 2019 & 0.165 & $0.254 \pm{0.031}$ & $2.122 \pm{0.020}$ & $-$0.04 \\
123215.16$+$132032.7 & Bright state & Graham et al. 2019 & 0.286 & $0.094 \pm{0.038}$ & $1.296 \pm{0.032}$ & $-$0.02 \\
123819.62$+$412420.5 & Bright state & Graham et al. 2019 & 0.499 & $0.000 \pm{0.011}$ & $1.592 \pm{0.069}$ & $+$1.01 \\
125757.23$+$322929.2 & Dim State & Graham et al. 2019 & 0.806 & $0.000 \pm{0.027}$ & $1.032 \pm{0.272}$ & $+$0.06 \\
132457.29$+$480241.2 & Bright state & MacLeod et al. 2016 & 0.272 & $0.304 \pm{0.107}$ & $1.184 \pm{0.044}$ & $+$1.11 \\
134822.31$+$245650.1 & Bright state & Graham et al. 2019 & 0.293 & $0.583 \pm{0.064}$ & $0.943 \pm{0.063}$ & $+$0.48 \\
143455.31$+$572345.0 & Bright state & MacLeod et al. 2018 & 0.175 & $0.094 \pm{0.032}$ & $1.086 \pm{0.089}$ & $+$1.65 \\
144202.82$+$433708.7 & Bright state & Graham et al. 2019 & 0.231 & $0.000 \pm{0.011}$ & $1.201 \pm{0.020}$ & $+$0.57 \\
144702.87$+$273746.7 & Bright state & Graham et al. 2019 & 0.224 & $0.000 \pm{0.015}$ & $1.694 \pm{0.015}$ & $+$1.39 \\
145022.73$+$102555.4 & Dim State & Graham et al. 2019 & 0.790 & $0.116 \pm{0.055}$ & $1.264 \pm{0.140}$ & $-$0.65 \\
145755.38$+$435035.4 & Dim State & Graham et al. 2019 & 0.528 & $0.243 \pm{0.096}$ & $1.325 \pm{0.101}$ & $-$0.17 \\
151604.24$+$355024.8 & Bright state & Graham et al. 2019 & 0.592 & $0.172 \pm{0.199}$ & $1.299 \pm{0.076}$ & $+$1.00 \\
153354.59$+$345504.1 & Bright state & Graham et al. 2019 & 0.753 & $0.000 \pm{0.011}$ & $1.456 \pm{0.172}$ & $+$0.15 \\
153612.80$+$034245.7 & Bright state & MacLeod et al. 2018 & 0.365 & $0.072 \pm{0.047}$ & $0.913 \pm{0.063}$ & $+$0.40 \\
153734.06$+$461358.9 & Bright state & MacLeod et al. 2018 & 0.378 & $0.540 \pm{0.166}$ & $0.823 \pm{0.203}$ & $+$1.09 \\
155651.38$+$321008.1 & Bright state & Graham et al. 2019 & 0.350 & $0.460 \pm{0.063}$ & $1.059 \pm{0.074}$ & $+$0.33 \\
160111.26$+$474509.6 & Bright state & MacLeod et al. 2018 & 0.297 & $0.157 \pm{0.071}$ & $0.982 \pm{0.067}$ & $+$1.48 \\
160742.94$+$432816.4 & Bright state & Graham et al. 2019 & 0.596 & $0.102 \pm{0.027}$ & $1.723 \pm{0.033}$ & $+$1.08 \\
161400.30$-$011006.0 & Bright state & Graham et al. 2019 & 0.253 & $0.219 \pm{0.055}$ & $1.175 \pm{0.062}$ & $+$0.71 \\
161711.42$+$063833.4 & Bright state & MacLeod et al. 2018 & 0.229 & $0.000 \pm{0.008}$ & $1.876 \pm{0.009}$ & $+$2.22 \\
162415.02$+$455130.0 & Bright state & MacLeod et al. 2018 & 0.481 & $0.517 \pm{0.128}$ & $1.268 \pm{0.057}$ & $+$1.14 \\
210200.42$+$000501.8 & Bright state & MacLeod et al. 2018 & 0.329 & $0.108 \pm{0.082}$ & $0.440 \pm{0.180}$ & $+$0.89 \\
214613.31$+$000930.8 & Dim State & MacLeod et al. 2016 & 0.621 & $0.000 \pm{0.012}$ & $1.757 \pm{0.054}$ & $-$0.22 \\
220537.71$-$071114.5 & Bright state & MacLeod et al. 2018 & 0.295 & $0.451 \pm{0.156}$ & $1.081 \pm{0.044}$ & $+$1.15 \\
224829.47$+$144418.0 & Bright state & Graham et al. 2019 & 0.424 & $0.319 \pm{0.096}$ & $0.754 \pm{0.082}$ & $+$1.44 \\
225240.37$+$010958.7 & Dim State & MacLeod et al. 2016 & 0.534 & $0.181 \pm{0.244}$ & $1.352 \pm{0.074}$ & $-$0.75 \\
231207.61$+$140213.3 & Dim State & Graham et al. 2019 & 0.357 & $0.074 \pm{0.025}$ & $1.459 \pm{0.062}$ & $-$0.91 \\
233136.83$-$105638.3 & Dim State & Graham et al. 2019 & 0.373 & $0.245 \pm{0.078}$ & $1.117 \pm{0.050}$ & $-$0.90 \\
233317.38$-$002303.4 & Dim State & MacLeod et al. 2016 & 0.513 & $0.316 \pm{0.308}$ & $1.524 \pm{0.064}$ & $-$1.36 \\
234307.38$+$003854.7 & Bright state & Yang et al. 2019 & 0.667 & $1.503 \pm{1.069}$ & $0.930 \pm{0.347}$ & $+$1.01 \\
235107.43$-$091318.0 & Bright state & MacLeod et al. 2018 & 0.355 & $0.051 \pm{0.057}$ & $0.942 \pm{0.066}$ & $+$1.25 \\
235439.14$+$005751.9 & Dim State & Graham et al. 2019 & 0.390 & $0.082 \pm{0.083}$ & $1.729 \pm{0.042}$& $+$0.53 \\
\end{longtable}

\subsection{Analysis 3: transition vectors on the EV1 plane of each object }

\begin{figure*}
\begin{center}
\includegraphics[width=180mm]{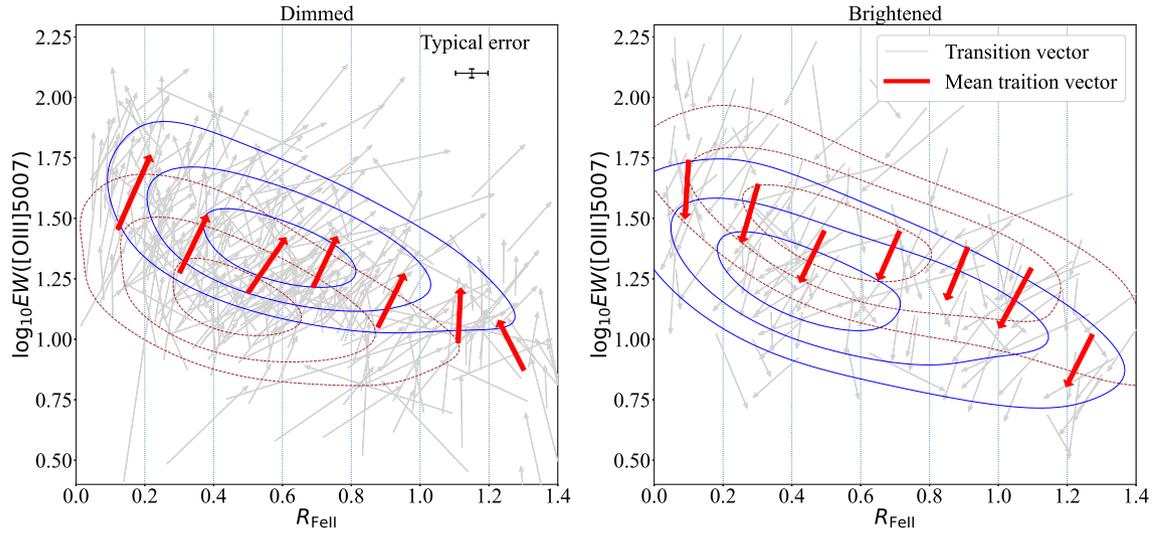}
\end{center}
\caption{
Transition vectors of quasars with multi-spectra on the EV1 plane. The left panel plots the dimmed sources, and the right panel plots the brightened sources. The positions on the EV1 plane of the newest and oldest spectra of the same source are calculated, and the same source is connected by a gray line. Only the spectra with the error less than $10\%$ of each emission line are used, and the mean value of the errors is indicated by a cross mark in the upper center of the left panel. The vertical dotted line divides the $R_{\rm{FeII}}$ into sections of 0.2 each. The red thick arrows show the average of the transition vectors of the objects whose starting points are in each of the seven regions separated by the dotted lines. Contours in blue solid brown dotted lines represent distribution after/before the variation using Gaussian kernel density estimation, respectively. \label{fig:arrow}
}

\end{figure*}

To understand how each quasar moves on the EV1 plane accompany with optical variation, we visualized transition vectors on the EV1 plane using Sample~3 (Figure~\ref{fig:arrow}). We fitted the spectra in this sample (8,142 sources) with PyQSOfit \citep{2018ascl.soft09008G} to determine their equivalent widths of emission lines. The wavelength range that we used for fitting is 4434-5535~$\rm{\AA}$ at the rest frame. The components applied in the fitting were power-law continuum, iron emission line templates \citep{1992ApJS...80..109B}, and two Gaussian components imposed for each $\rm H\beta$, ${\rm [OIII]5007}$, and ${\rm [OIII]4959}$. The velocity offsets and the line width for the narrow emission lines (${\rm [OIII]5007}$, ${\rm [OIII]4959}$) were set to be equal. Each measurement error was estimated using the Monte Carlo method, which perturbs the flux based on the input error and the fitting was repeated 20 times. The equivalent widths were calculated from the fitting results. The wavelength range of $\rm{FeII}$ emission lines is 4435-4685~$\rm{\AA}$ at rest frame in this calculation. We narrowed down the list to 2,839 sources for which all the measured emission lines in the multiple spectra were obtained with an error of less than 10\%. 
Figure~\ref{fig:arrow} shows a graph of the transition vectors of each source on the EV1 plane, connected by straight gray arrows. From the fitting results, we classified the variability of the sources on the basis of the assumption that the ${\rm [OIII]5007}$ luminosity is constant. Among these sources, those with $\Delta {\rm log_{10}} EW({\rm [OIII]5007})$ greater than 0.15 are grouped as ``dimming sources'' (325 sources), and those with $\Delta {\rm log_{10}} EW({\rm [OIII]5007})$ less than -0.15 are grouped as ``brightening sources'' (156 sources). 
The difference in the periods of observations of the spectra being compared is typically about eight years in the rest frame (the median value is 3,024 days, and $80\%$ is longer than 1,500 days).
Based on the initial (the first observation in multiple spectra) $R_{\rm{FeII}}$ values, the sources are divided into seven groups, whose averaged transition vectors are represented by red thick arrows. 

\begin{figure*}[ht!]
\includegraphics[width=160mm]{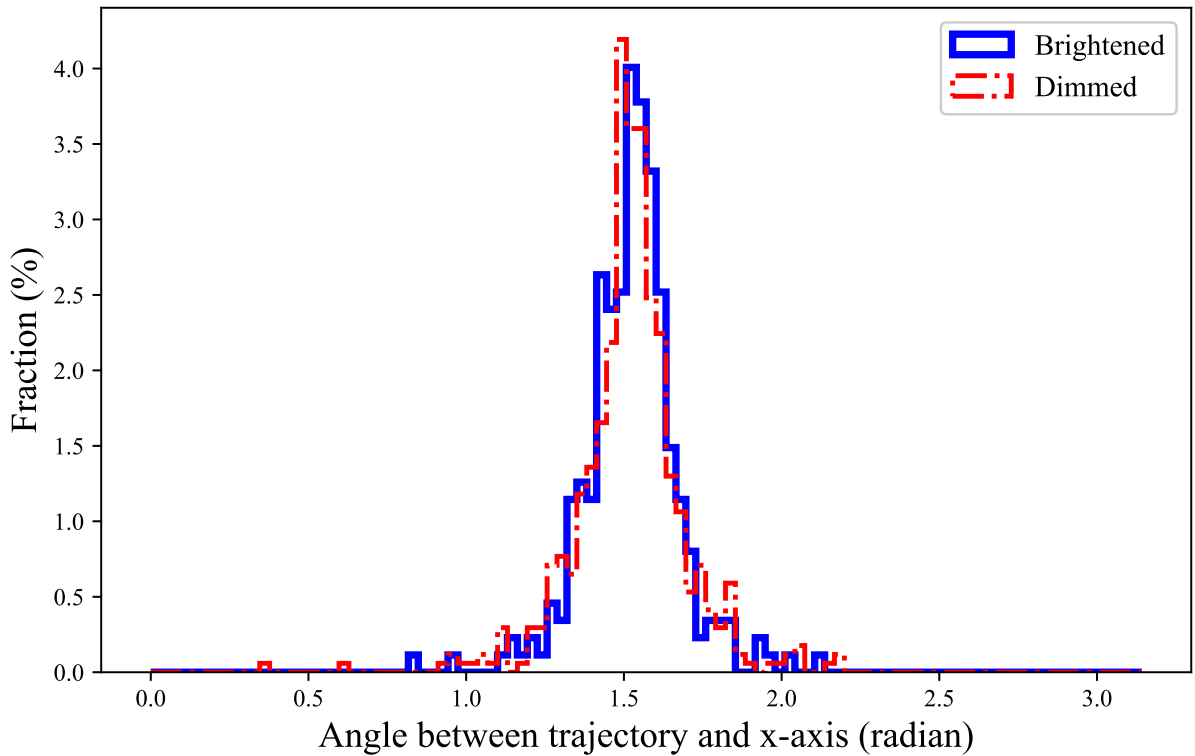}
\caption{
Probability distribution of the angle between the transition vectors and the x-axis expressed in radians. The blue line represents the brightened sources, and the red dashed-dotted line represents the dimming sources.}
\label{fig:hist}
\end{figure*}

As a result, we can see that both the brightening and dimming sources follow a similar path and swap positions on the EV1 plane. They go back and forth between the lower left and upper right on the EV1 plane, each position is expected to correspond to a bright state and a dim state from Analysis~2. To compare the transition vectors of brightening and dimming, we illustrate a histogram of the inclination of the transition vectors (Figure~\ref{fig:hist}). In order to quantitatively verify the two distributions in Figure~\ref{fig:hist}, we performed a two-component Kolmogorov-Smirnov test. The p-value is 0.83, indicating that the Null hypothesis that these two distributions are the same is not rejected. This result confirms that quasars move in almost the same direction when they brighten or dim, and they transit to the opposite state.

We also note here that when each quasar brightens, $R_{\rm{FeII}}$ decreases and vice versa. This means that $R_{\rm{FeII}}$ is anti-correlated with the Eddington-ratio if we focus on each quasar. This anti-correlation is opposite to the general trend derived from the statistical analysis of single epoch spectra of quasars \citep{1992ApJS...80..109B, 2000ApJ...536L...5S, 2014Natur.513..210S}. 

\section{Discussion} \label{sec:discussion}
This study explored the relationship between optical variability and spectra through three analyses. The first analysis used the catalog values of S11 to visualize the distribution on the EV1 plane with the subsequent variability. Second, we used a sample of CLQs to show the distribution on the EV1 plane depending on their bright or dim status. Third, we visualized the transition of the positional transition of multi-epoch spectra on the EV1 plane. We summarize the main results of these analyses below.

\begin{itemize}
\item Tendency of subsequent brightness variation is related to the position on the EV1 plane. Sources that dim later distribute on the lower side of EV1, and sources that brighten later distribute upper side (Analysis 1; Figure~\ref{fig:shen_variation}). 
\item In general, significantly variable quasars distribute on the left side of the distribution ridge on the EV1 plane. (Analysis 1; Figure~\ref{fig:shen_variation})
\item In the case of CLQs, bright/dim state objects are located at the lower-left/upper-right side of the distribution ridge on the EV1 plane, respectively (Analysis 2; Figure~\ref{fig:clq_ev1}).
\item On the EV1 plane, the brightening and dimming sources cross the ridge of the general distribution and follow similar transition vectors, moving back and forth between the lower left and upper right. (Analysis 3; Figure~\ref{fig:arrow})
\item When we focus on an individual object with multi-epoch spectra, $R_{\rm{FeII}}$ is anti-correlated to the Eddington ratio. (Analysis 3; Figure~\ref{fig:arrow} and Figure~\ref{fig:hist})
\end{itemize}

These are new results that confirm link quasars' random variation with spectral features. Based on them, we will discuss from four viewpoints in the following subsections; distribution on the EV1 plane, transition vectors on the EV1 plane, limitations of the present method, and observational predictions.

\subsection{Distribution on the EV1 plane}
\label{sec:distribution_ev1}
Since EV1 is a relation found phenomenologically using principal component analysis \citep{1992ApJS...80..109B}, it is still poorly understood physically. Therefore, this section discusses the meaning of each of the values of $EW({\rm [OIII]5007})$ and $R_{\rm{FeII}}$ that constitutes EV1.

${\rm [OIII]5007}$ is a narrow emission line commonly observed in quasars. Narrow lines are usually powered by ionization photons from the accretion disk, and thus are affected by the energy distribution of the disk. However, the timescale for the change in ${\rm [OIII]5007}$ brightness is considered to be sufficiently more prolonged than the observation timescale, because the ${\rm [OIII]5007}$ emission region is substantially extended ($\sim$100 pc from the center). In other words, this region reflects the history of changes in quasar activity over the past $\sim$100 years. Since we are interested in the amount of $EW({\rm [OIII]5007})$ changes over the past $\sim$10 years, we interpret that the change in $EW({\rm [OIII]5007})$ is dominated by the continuum variation.

$R_{\rm{FeII}}$ has been considered to be a positively correlated indicator of mass accretion rate based on the trends of many objects \citep{1992ApJS...80..109B, 2000ApJ...536L...5S, 2014Natur.513..210S}. The result of Analysis~3, however, shows that a negative correlation exists in the variation of individual quasar, with $R_{\rm{FeII}}$ decreasing as brightening. Some reverberation mapping studies have confirmed that the $\rm{FeII}$ emission region is more extended than the $\rm{H\beta}$ region \citep{2013ApJ...769..128B, 2019ApJ...876...49Z, 2020ApJ...905...75H, 2021ApJ...918...50L, 2022AN....34310112G}. This may cause slower and smoother variation of $\rm{FeII}$ emission lines compared to that of $\rm{H\beta}$, which makes $R_{\rm{FeII}}$ temporarily smaller/larger during brightening/dimming, respectively. 

How much extend $\rm{FeII}$ emission region is under debate, but it is likely comparable to the dust sublimation radius  (e.g., \cite{2021ApJ...907L..29H, 2021ApJ...913..146M, 2022AN....34310112G}). The typical sublimation radius is $\sim100$ light days \citep{2016MNRAS.460..980N}, but the typical difference in observation epoch in Analysis~3 is $\sim1,000$ days. Thus, In the case of Analysis~3, only the difference in the distributed radius of $\rm H\beta$ and $\rm{FeII}$ is not the dominant explanation for the negative correlation, because the light crossing time of the $\rm{FeII}$ emission region is significantly shorter than the observation timescale. 

Although it is commonly accepted that $R_{\rm{FeII}}$ has a correlation with the Eddington ratio, the Pearson correlation coefficient between log(Edd) and log($R_{\rm{FeII}}$) is as low as 0.25 for the Sample~1. 
The scatter in the distribution may have been increased by the differences in the distribution of $\rm H\beta$ and $\rm{FeII}$, the selection method of Analysis~3, or other factors related to the origin of EV1 may have increased.
Further research is needed to clarify how much they are correlated as well as the effect of the temporal variation of $R_{\rm{FeII}}$ on it. High precision reverberation mapping will identify the location of the ${\rm FeII}$ emission region, one of the clues to the actual physical meaning of $R_{\rm{FeII}}$.

On the other hand, the overall EV1 of S11, $EW({\rm [OIII]5007})$ and $R_{\rm{FeII}}$ are weakly correlated (Pearson's correlation coefficient is -0.35). The major factors that determine this correlation of emission line ratio are elemental composition, electron density, electron temperature, and energy distribution of ionizing photons. In timescale of some years, the key factor that contributes to the variation is the energy distribution of ionizing photons, i.e., the accretion state characterized by the Eddington ratio. When we interpret that the whole EV1 correlation is produced by a certain systematic variation of the SED of each source according to the Eddington-ratio change (modifying the structure of the accretion flow) as suggested in \cite{2014Natur.513..210S}, the vertical scatter along the ridge of the distribution on the EV1 plane reflects the current level of mass accretion activity normalized by the averaged past activity over $\sim 100$ years (the timescale of ${\rm [OIII]5007}$ variation).

We also mention intrinsic scatter contribution around the ridge on the EV1 plane. The intrinsic scatter (including a variation on timescales sufficiently longer than observations, opening angle of the UV continuum radiation, the profile of UV continuum, metallicity, and the possible contribution of  $\rm Ly\alpha$ pumping (\cite{2021ApJ...907...12S}, and references therein)) and the short-term variation are expected to contribute to the spread of the distribution on the EV1 plane. The vertical scatter of ${\rm log_{10}}EW({\rm [OIII]5007)}$ on the EV1 plane of Sample~1 quasars has a standard deviation of 0.28. Assuming that the luminosity of ${\rm [OIII]5007}$ does not change during observations, a 0.28 change in ${\rm log_{10}}EW({\rm [OIII]5007})$ means that the continuum changes by about 0.71 mag. On the other hand, the structure function of 20 years light curves indicates that the standard deviation of the magnitude difference between the data with 10 years separation is about 0.3 mag \citep{2022MNRAS.514..164S}. This comparison indicates that the scatter of the distribution on the EV1 plane is more extensive than would be explained by the typical continuous light variation over ten years.

In summary, though intrinsic scatter exists, we interpret that the location on the EV1 plane is significantly affected by each quasar's activity level normalized by the average of $\sim 100$ years. The more out of the peak of the distribution on the EV1 plane, the more deviate the activity state is, compared to the typical state for the object. In other words, quasars at the lower on the EV1 plane are in temporarily bright state, while those at the higher on the EV1 plane are in temporarily dim state.

\subsection{Transition vectors on the EV1 plane}

Figure~\ref{fig:shen_variation} shows a clear difference in subsequent brightness variation depending on the locations on the EV1 plane. From Figure~\ref{fig:clq_ev1} and Figure~\ref{fig:arrow}, we can see that the lower left side of the EV1 plane corresponds to the quiescent phase for the source, while the upper right side corresponds to the active phase. Figure~\ref{fig:arrow} shows the averaged transition vectors located across the ridge of the distribution, from the quiescent (lower-left) region to the active (upper-right) region, and vice versa. Furthermore, the transitions on the EV1 plane follow the same path as the results of Figure~\ref{fig:arrow} and Figure~\ref{fig:hist}. From these results, interpreting the distribution on the EV1 plane as the current activity level normalized by the activity level over the past several hundred years (the timescale of ${\rm [OIII]5007}$ changes), it is suggested that the activity level transits between the quiescent and active phases across the equilibrium state for each object.

The periodic accretion disk fluctuations are similar to the limit cycle caused by the viscous and thermal instability of the accretion disk. Based on the standard disk models consistent with the SEDs of many AGNs, thermal instability is known to occur in situations where the radiation pressure of the accretion disk is more dominant than the gas pressure (e.g., \cite{1986ApJ...305...28L}). In thermally unstable conditions, the radius of the 50-150$R_g$ ($R_g$ is the gravitational radius) of the accretion disk is known to transit between a state composed of ionized hydrogen and a state composed of neutral hydrogen, which has been confirmed in dwarf novae and X-ray binary stars. It has also been shown that state transitions can occur in the accretion disks of AGN, making it one of the major causes of CLQs (e.g., \cite{2018MNRAS.480.3898N}). The disk instability model predicts the existence of a limit cycle, which has also been proposed for AGNs \citep{1986ApJ...305...28L}. Moreover, the transition timescale of CLQs ($\sim$ several years) is comparable with that of the disk instability ($\sim10$ years).

Here, we propose one possibility that the evolutionary process of AGN is able to be inferred from their distribution on the EV1 plane. Since there is a time lag of $\sim 100$ years between the ${\rm [OIII]5007}$ luminosity and the central activity, we can infer the history of quasars' activity using ${\rm [OIII]5007}$. For example, when a quasar ends its activity in $\sim 100$ years, the ${\rm [OIII]5007}$ luminosity is expected to be large compared to the central core \citep{2019ApJ...870...65I}. If we assume quasars generally fade to their end of activity (\cite{2020ApJ...889L..29C} observationally confirmed that most quasars are fading), the central luminosity is expected to generally decrease in the long-term ($\sim 100$ years) in addition to short-term (months to years) temporal variation. Under this assumption, quasars' central luminosity compared to ${\rm [OIII]5007}$ will also be decreased because there is $\sim 100$ years of time lag between them. That is to say, if a quasar is close to the end of its activity, the $EW({\rm [OIII]5007})$ is expected to be larger (the upward on the EV1 plane). Now, it has been known that objects with small black hole masses and large Eddington ratios, such as Narrow Line Seyfert 1, are characterized by large $R_{\rm{FeII}}$, while objects with large black hole masses and small Eddington ratios are characterized by small $R_{\rm{FeII}}$ \citep{2018FrASS...5....6M}. When we interpret this fact as that the quasars with larger $R_{\rm{FeII}}$ are in the early stage of an actively growing AGN, and the smaller $R_{\rm{FeII}}$ means the late stage of the growth, which can explain the negative correlation of the EV1 plane. Also, such an interpretation can explain that the significantly variable sources are mostly distributed on the left side of the EV1 plane (Fig~\ref{fig:shen_variation}). Objects on the upper left (larger $EW({\rm [OIII]5007})$) are less efficient at mass accretion and more prone to intermittent mass accretion like an avalanche \citep{1995PASJ...47..617T, 1998ApJ...504..671K, 2020ApJ...903...54T} because they are in a quiescent phase compared to past AGN activity. In summary, our hypothesis is that the sources distributed in the lower right on the EV1 plane are in the active growth phase of AGN, while those in the upper left are in the terminal phase of AGN activity.

To summarize what we found from the quasars' trajectories on the EV1 plane, if the ridge of the distribution on the EV1 plane corresponds to averaged activity level (as mentioned in Section~\ref{sec:distribution_ev1}), the status variation across the ridge are expected to oscillate around the ridge. In fact, quasars with large brightness variation generally transit across the ridge along the similar path during both brightening and dimming. Based on these pieces of circumstantial evidence, it seems that quasars' large variations, such as CLQs, are generally repeating phenomena.

\subsection{observational predictions}
If quasars' significant variation is due to disk instability of the accretion disk occurring repetitively, three things are expected to be observationally confirmed.

\begin{itemize}

\item Since the timescale inferred from the disk instability model is shorter for brightening and longer for dimming as confirmed in dwarf novae \citep{1996PASP..108...39O}, it is expected that the CLQs discovered in the future will be more likely to be dimming sources regardless of the selection bias.

\item The known CLQs are more likely to transit activity state later than typical quasars. We suggest that known CLQs are good targets to follow up to investigate the state transition process observationally.

\item The distribution of the dimming/brightening CLQs in the EV1 plane (i.e., Figure~\ref{fig:clq_ev1}) will be further enhanced by newly discovered CLQs in the future. This difference in distribution can be applied to the sample selection when searching for new CLQs. This suggests that the subsequent variability of quasars can be predicted from their spectra.

\end{itemize}

The above observational predictions can be verified in the future with the development of survey observations such as The Large Synoptic Survey Telescope \citep{2009arXiv0912.0201L}. On the contrary, the results obtained in this study should be treated with a certain degree of caution, because they were obtained from a sample with a non-uniform observation period. In the next section, we discuss the limitations of this study.

\subsection{Limit of this method and Future work}
In this paper, we showed the relation between optical variation and spectra. Our analysis tried to use as large samples as possible to reduce selection bias of the sample. However, we understand that the method in this paper has two shortcomings, which we should keep in mind.

The first is that the timing of the spectroscopic and photometric observations is determined by chance, making the intervals between observations not uniform. Although the Pan-STARRS value is used as the magnitude of the object about 10 years later than the SDSS, the interval between the two is actually uncertain for several years. It should be noted that the results of this paper only claim to reveal an overall qualitative trend by using a large sample.

Second, we still have a limited understanding of EV1. Although the correlation was discovered by a phenomenological approach among various physical quantities, the mechanism seems not simple as shown in this study. Further investigation is required for this issue.

It is expected that the above two problems will be improved by future research. The first problem is expected to be solved by revalidating the data on the basis of homogenized data accumulated in the future. The second problem is expected to be solved by clarifying the elemental (FeII, in particular) distribution by reverberation mapping and interpreting the trend on EV1 from the physical model.

\section{Conclusion} \label{sec:conclusion}

In this study, we visualized the relationship between optical variation and location on the EV1 plane. We summarize what we found and what can be inferred from the results below.
\begin{itemize}
\item There is a correlation between quasars' optical variability and their distribution on the EV1 plane, in which the significantly dimming objects tend to transit across the distribution ridge from lower-left to upper-right, and vice versa for the brightening objects.
\item $R_{\rm{FeII}}$ is anti correlated to the Eddington-ratio when we focus on indivisual quasar.
\item Reported CLQs are expected to repeat state transition.
\end{itemize}

\begin{ack}
This work is supported by JSPS KAKENHI Grant Number JP22J13428. The data presented in this paper were obtained from Sloan Digital Sky Survey.
Funding for the Sloan Digital Sky Survey IV has been provided by the Alfred P. Sloan Foundation, the U.S. Department of Energy Office of Science, and the Participating Institutions. SDSS-IV acknowledges support and resources from the Center for High-Performance Computing at the University of Utah. The SDSS web site is www.sdss.org.
SDSS-IV is managed by the Astrophysical Research Consortium for the 
Participating Institutions of the SDSS Collaboration including the 
Brazilian Participation Group, the Carnegie Institution for Science, 
Carnegie Mellon University, the Chilean Participation Group, the French Participation Group, Harvard-Smithsonian Center for Astrophysics, 
Instituto de Astrof\'isica de Canarias, The Johns Hopkins University, Kavli Institute for the Physics and Mathematics of the Universe (IPMU) / 
University of Tokyo, the Korean Participation Group, Lawrence Berkeley National Laboratory, 
Leibniz Institut f\"ur Astrophysik Potsdam (AIP), 
Max-Planck-Institut f\"ur Astronomie (MPIA Heidelberg), 
Max-Planck-Institut f\"ur Astrophysik (MPA Garching), 
Max-Planck-Institut f\"ur Extraterrestrische Physik (MPE), 
National Astronomical Observatories of China, New Mexico State University, 
New York University, University of Notre Dame, 
Observat\'ario Nacional / MCTI, The Ohio State University, 
Pennsylvania State University, Shanghai Astronomical Observatory, 
United Kingdom Participation Group,
Universidad Nacional Aut\'onoma de M\'exico, University of Arizona, 
University of Colorado Boulder, University of Oxford, University of Portsmouth, 
University of Utah, University of Virginia, University of Washington, University of Wisconsin, 
Vanderbilt University, and Yale University.
\end{ack}
\bibliography{main}{}
\bibliographystyle{pasj_bibtex.bst}

\end{document}